# Memory Testing Under Different Stress Conditions: An Industrial Evaluation

Ananta K. Majhi, Mohamed Azimane,
Guido Gronthoud, and Maurice Lousberg
Philips Research Laboratory
Eindhoven, The Netherlands

Stefan Eichenberger
Philips Semiconductors
Nijmegen, The Netherlands

Fred Bowen
Philips Semiconductor
San Jose, CA-95131

*Abstract*

*This paper presents the effectiveness of various stress conditions (mainly voltage and frequency) on detecting the resistive shorts and open defects in deep sub-micron embedded memories in an industrial environment. Simulation studies on very-low voltage, high voltage and at-speed testing show the need of the stress conditions for high quality products; i.e., low defect-per-million (DPM) level, which is driving the semiconductor market today. The above test conditions have been validated to screen out bad devices on real silicon (a test-chip) built on CMOS 0.18 um technology. IFA (inductive fault analysis) based simulation technique leads to an efficient fault coverage and DPM estimator, which helps the customers upfront to make decisions on test algorithm implementations under different stress conditions in order to reduce the number of test escapes.*

## 1. Introduction

What is the yield of your SoC (System-on-Chip)? If one can give the yield of embedded memories in the SoC, then it becomes very easy to predict the SoC yield. The simple fact is that nowadays SoCs are becoming very much memory dominant (71% by 2005 – ITRS 2001). Additionally, the embedded memory size is also increasing day by day, e.g., 256Mbits and more. The feature size and large chip area result in an enormous critical area for defects. Due to the fact that memory yield is decreasing because of increasing size, the overall SoC yield may be unacceptable, unless special measures have been taken [Hamdioui 04]. Growing gate counts, change in process technology and increasing soft defects is a real test challenge for keeping the quality level (e.g., ~10 DPM for automotive market) under control.

Traditionally, march tests are used for testing semiconductor memories and target the classical fault models with the emphasis on faults in the matrix [vdGoor 98, Adams 02, Borri 03]. These functional fault models employed in memory testing, such as stuck-at, transition and coupling faults have become insufficient to properly model the effects of the real defects occurring in deep sub-micron memories. Resistive shorts and opens are types of defects (also termed as soft defects) that require specific test pattern sequences under particular stress conditions for detection. Resistive shorts were quite predominant in CMOS 0.18um technology and above. However, open and resistive open defects are becoming more dominant as the technology is moving from aluminum to copper interconnects in more advanced technologies (like CMOS 0.13um and beyond).

Industrial researchers have shown that open and resistive vias are the main root cause for test escapes in deep sub-micron technologies [Needham 98]. These types of defects mainly lead to timing dependent behavior that can only be detected by at-speed testing using a special test sequence. One such defect due to a salicide break occurred inside a flip-flop (shown in Figure 1) in a CMOS 0.13um process and was detected *only* by applying delay test patterns [Kruseman 04].

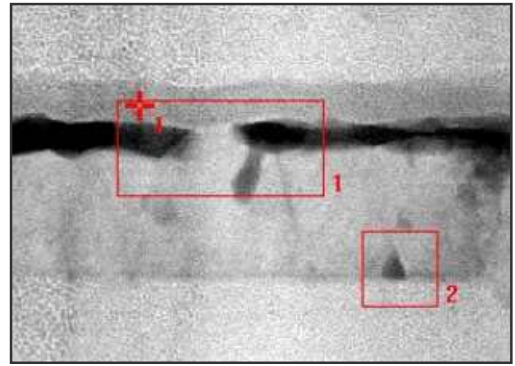

*Figure 1*: Salicide break leads to resistive-open defect behavior in a CMOS 0.13um technology

The primary target of this study is to analyze the defect behavior of resistive shorts and opens in sub-micron memories and the test conditions for detecting these soft defects in order to achieve an acceptable DPM level. For both types of defects, very extensive analogue simulations have been performed employing various stress conditions such as defect resistance, supply voltage, timing and defect density in the process. The outcome of this investigation leads to a novel fault coverage and DPM estimator, which is a powerful tool for our customers for delivering high quality products. We have also conducted an experimental study on real silicon on a test chip fabricated in CMOS 0.18um process as proof of concept.

The paper is organized as follows: Section 2 describes the simulation and experimental setup. Fault coverage and DPM estimator is given in Section 3. The stress conditions are highlighted in Section 4 while Section 5 gives the experimental summary. Finally, Section 6 concludes the paper with recommendation for high quality testing.

## 2. Simulation and Experimental Setup

The simulation flow is based on the ideas behind inductive fault analysis (IFA) [Shen 85]. For targeting resistive shorts in our simulation flow, bridging faults are extracted from the layout structure. The bridge extraction is based on the critical area of two neighboring nodes and is performed by setting the extraction rules for a determined technology. The flat fault-free netlist is extracted from the same layout using Philips internal tool (PIA). The extracted bridges are added one-by-one to the fault-free netlist in order to get the faulty netlist, which is then simulated using a Spice-like simulator. One defect at a time is injected into the extracted fault-free netlist. The analogue input stimulus with respect to the family of march tests have been automated to create the test bench needed for simulation. The pictorial presentation of the simulation flow is given in Figure 2. The simulation process is almost the same for analyzing the resistive open defects. In the following section, we have presented the simulation results.

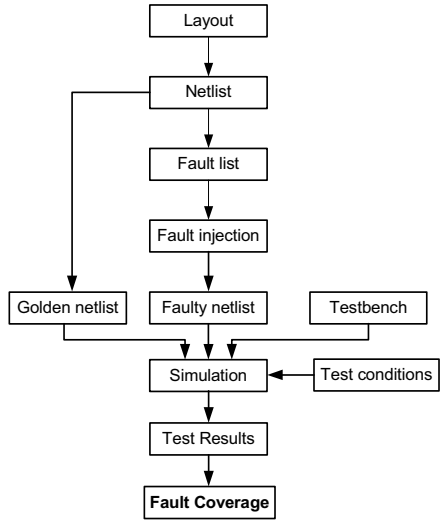

*Figure 2*: Simulation flow

In order to validate our simulation results, we conducted an experiment on silicon in which a few thousands of SRAMs were assembled and tested by applying test algorithms under different test conditions. For this investigation, we have presented the result of the 11N March test, which is a variation of MATS++, March C- and MOVI. The test chip (Veqtor4; built on CMOS 0.18um technology) contains four instances of SRAMs of 256 K bits each. Each of the memory cores can be accessed directly from the primary inputs/outputs through a controller. Memory BIST was not implemented at the time of design as this test chip was only intended for process qualification.

## 3. Fault Coverage and DPM Estimator

Calculating the fault coverage precisely would take years of simulation time, but using a database with pre-calculated simulation results makes the fault coverage estimation an easy job. We have set up a memory test flow based on IFA analysis to generate a database from which the fault coverage is calculated. The memory test flow results are collected in a database. The users can enter the four design parameters to the Fault Coverage Estimator which are: the #X rows (number of Xrows), the #Ycolumns (number of Ycolumns), the #B (number of bits per word) and the number of Z blocks (optional). The estimator gives the fault coverage and the DPM level based on a certain yield. We relieve the users from the burden of running a time consuming IFA analysis.

Table 1 summarizes the fault coverage estimation results with respect to different supply voltage and defect resistance. This table highlights the results of only resistive bridges in CMOS 0.18um technology. We have chosen four supply voltage conditions - nominal of 1.8V, Vdd min/max (+/- 10% of nominal), and very-low voltage of 1.0V. For low ohmic bridges (~20 ohm), all four-supply voltage conditions give reasonably good fault coverage (above 95%). However, for high-ohmic bridges (~90 Kohm), Vnom and Vmax give very low fault coverage, whereas very-low voltage testing shows much higher coverage. This is an indication of importance of very-low voltage testing needed to detect high-ohmic resistive bridge defects. Similar investigations show that high voltage testing is necessary to get higher fault coverage in resistive open defects. This same observation has been made recently [Borri 03].

### 3.1 Defect Coverage and DPM estimation

Defect coverage is different than that of fault coverage as shown in Table 1. The distribution of the defect resistance is obtained from the fab, combining that with the weighted fault coverage, we obtain a more realistic defect coverage of resistive bridges as given in the table.

The DPM is calculated based on Williams and Brown model [Williams 81];
$$DPM = 1 - Y^{(1-DC)} \qquad (1)$$
where DC is the defect coverage and Y is the yield.

| Test Conditions: Vdd | Test Voltage (V) | Fault Coverage by bridge defect resistance | | | | Defect Coverage | DPM |
|---|---|---|---|---|---|---|---|
| | | 20 ohm | 1 Kohm | 10 Kohm | 90 Kohm | | |
| | 1.00 – VLV | 99.61 | 98.57 | 98.57 | 88.90 | 98.92 | 1x |
| | 1.65 – Vmin | 97.76 | 86.95 | 86.95 | 77.91 | 95.15 | 4.4x |
| | 1.80 – Vnom | 97.58 | 87.90 | 86.95 | 30.81 | 95.10 | 4.45x |
| | 1.95 - Vmax | 95.65 | 87.89 | 87.82 | 1.22 | 89.76 | 9.3x |

*Table 1*: Defect Coverage and DPM Estimator

The yield is calculated based on the following formula:
$$Y = e^{-AD0} \quad (2)$$
where A is the area of the chip and D0 is the defect density obtained from fab data.

It is clear from the DPM estimation (Table 1) that very-low voltage testing is unavoidable in order to reduce test escapes. We have normalized the DPM level for VLV testing to 1x. Also, it is to be noted that there is almost an order of magnitude difference in the DPM level when we compare between VLV and Vmax testing (e.g. 9.3 x). Interestingly, we obtained almost identical results from our experiments on silicon. The following sections will highlight the various test conditions and experimental validation.

## 4. Stress Conditions

Importance of very-low voltage testing is not new at this moment. Many researchers have explored the concept of very-low voltage testing in the last decade [Chang 96, Schantra 99, Li 01, Kruseman 02, Engelke 04]. However, the importance of high voltage testing is slowly becoming popular, in particular when targeting resistive open defects [Borri 03]. At-speed testing intended to target mainly timing related failures (known as delay faults) caused by resistive bridges as well as open defects. In this section, we will elaborate the above test conditions in more detail with the experimental validation.

### 4.1 Testing @ VLV

The reduction in supply voltage (Vdd) decreases the driver strength of the transistors and hence makes it more sensitive to resistive shorts [Chang 96]. Earlier simulation [Kruseman 02] also has shown that with a reduced supply voltage of 1.5 $V_T$, one can detect shorts with five times higher resistance than can be detected at nominal voltage (4 $V_T$). We have also shown in Section 3.1 the defect coverage of known march test versus the supply voltage. Table 1 highlights the importance of VLV testing for better coverage of resistive bridges. In order to pass the test at VLV test conditions, one must decrease the test frequency. Hence test time can become an issue in a production environment. However, one must consider the trade-off to achieve high quality products. Additionally, the device must be characterized in order to determine the optimum test frequency and VLV level. The test engineer can readily accomplish this task.

Figure 3 shows a tester-generated shmoo plot (voltage in Y-axis vs. clock period in X-axis) for a fault-free chip. The horizontal dashed lines show the nominal and VLV test conditions, whereas the vertical dashed line is the clock period at which the memories are tested. In order to show the effectiveness of VLV, we are testing our memories at slow frequency of 10MHz (100ns clock cycle).

Normally, the chips are tested at Vnom (1.8V), Vmin (1.65V) and Vmax (1.95V). However, for our investigation on the effectiveness of VLV testing, we added this new stress condition to test the memories at 1.0V. This very-low voltage value is well within the window limit of 2 to 2.5 $V_T$ as recommended by earlier researchers [Chang 96, Kruseman 02]. The point to be highlighted here is that the fault-free SRAM still passes at 100ns clock period under VLV testing as shown in Figure 3.

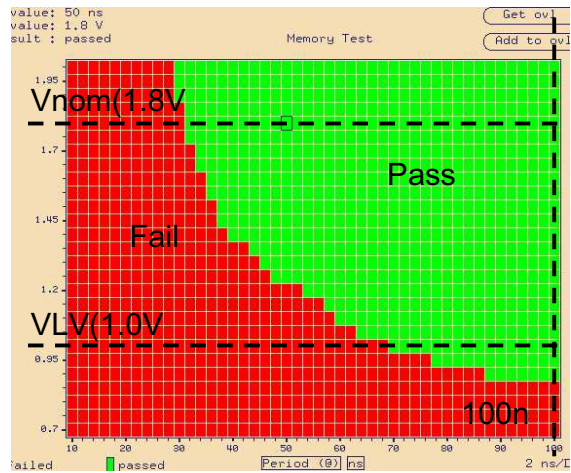

*Figure 3*: Shmoo plot (Vdd vs. period) for a fault-free SRAM (as reference)

The shmoo plot for a faulty device (Chip-1) is shown in Figure 4. Chip-1 would have been declared as fault-free if tested under normal test conditions (Vmin, Vnom and Vmax @ 100ns). However, it was only possible to mark this device as a faulty one by applying the same test patterns at very-low voltage. The lower horizontal dashed line clearly shows that the device failed at VLV test condition as shown in Figure 4.

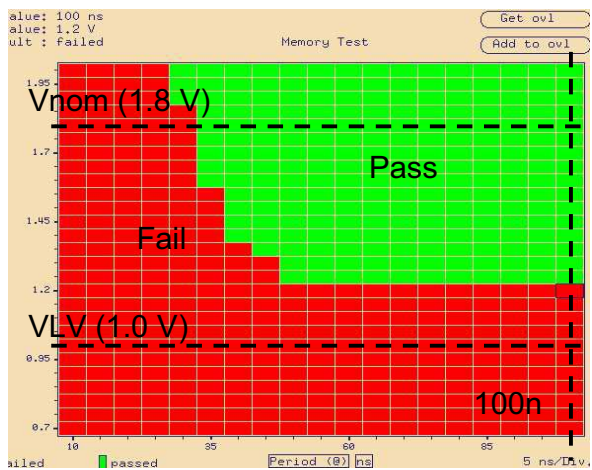

*Figure 4*: Shmoo plot for Chip-1 (Fails at 1.0 V/100ns)

The bitmapping result shows the failure in three clock cycles that belong to three march elements of the applied test; they are [i] {$R_0W_1$}, [ii] {$R_1W_0R_0$}, and [iii] {$R_0W_1R_1$}. In all cases, this points to the same address location/cell and fails while reading '0' at the same output. Hence, we conclude that there could be a resistive bridge, which is leading to a stuck-at-1 behavior in a single cell *only* at lower supply voltage. The resistive bridge which may behaving as a voltage divider is not sensitive enough at higher voltages (>1.2V) as shown in Figure 4 in order to give faulty outputs.

**4.2 Testing @ Vmax**

Similarly, in order to investigate the best stress conditions to detect resistive open defects we have carried out an experiment using analogue simulation and validated the same on silicon. IFA analysis is used to extract open defects from the layout, which have been simulated to compare the different stress conditions. Figure 5 shows the simulation result of an open defect injected at the least significant bit of the row address decoder. Our standard test patterns have been applied at normal test frequency, room temperature and at nominal voltage. The injected open defect escaped our test at these test conditions. We then simulated the same faulty netlist under the VLV test conditions, and again the defect escaped the test.

the failure in two clock cycles that belong to two march elements of the applied test; they are [i] $\{R_0W_1\}$, and [ii] $\{R_0W_1R_1\}$. In both cases, this points to the same address location / cell and fails while reading '0'. Thus, it is also a single bit failure in the matrix.

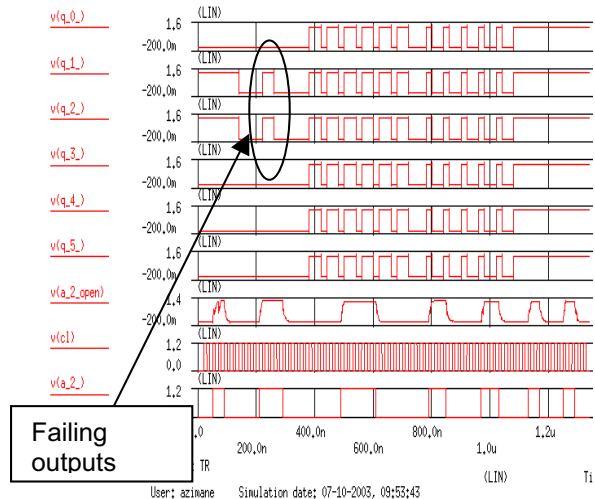

*Figure 6*: Simulation results of the same open defect, detected at Vmax test

Both experiments (i.e. analogue simulation and test of real devices) have shown the need of high voltage testing to detect resistive open defects in semiconductor memories. No other stress conditions or specific test patterns were found to detect all open defects. Therefore, the need for high voltage testing to detect resistive open defects is essential.

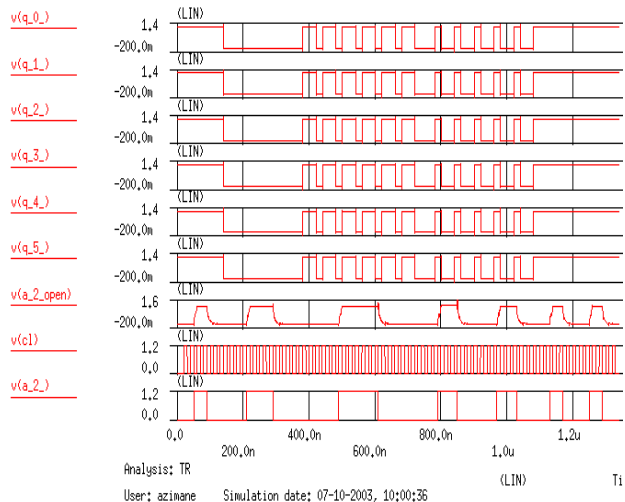

*Figure 5*: Simulation results of an injected open defect at the memory address decoder (case of Vnom)

The same open defect has been detected by the same test at higher supply voltage (Vmax). Figure 6 shows the example of the simulation results in which the defect has been detected during a unique clock cycle at the memory outputs **q1** and **q2**. This simulation proves that Vmax test forces the resistive open defect to propagate the faulty behavior to the memory outputs. We previously showed that Vnom and VLV tests were not able to sensitize and propagate the defect. Thus Vmax offers a unique stress condition, which helps to increase the fault coverage of resistive open defects.

The above analysis has been validated through experiment on silicon where certain devices fail only the Vmax test. Figure 7 shows a shmoo plot for Chip-2 that fails only the Vmax test and passes both Vnom and VLV tests irrespective of test frequency. The bitmapping result shows

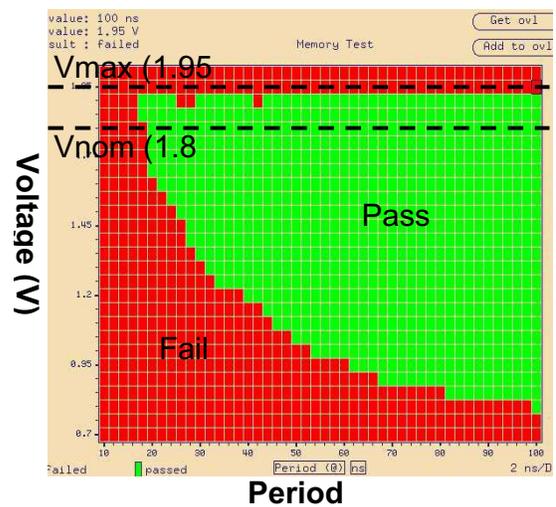

*Figure 7*: Shmoo plot for Chip-2 (Fails only at Vmax and above)

### 4.3 Testing @ at-speed

At-speed testing is necessary to target timing related failures also known as dynamic faults [Borri 03, Azimane 04]. However, at-speed testing does not mean running the standard test pattern at high speed, instead it requires a dedicated test pattern sequence be applied at the highest specified speed of the memory to increase the fault coverage of resistive defects. We have investigated the correlation between at-speed test and resistive open defects by using defect oriented test simulation. The result shows that a particular test frequency detects only a specific range of resistive open defects. For instance, testing at 50 MHz a memory that operates at 100MHz will detect resistive open defects above 4MΩ as shown in Figure 8. But, all resistive open defects below 4MΩ will escape the test. In order to reduce test escapes, the test must be run at 100 MHz. However, even at that speed, resistive open defects below 1.5MΩ still escape the test. Hence, it is recommended to test at even relatively higher frequency than the specified speed to reduce test escapes.

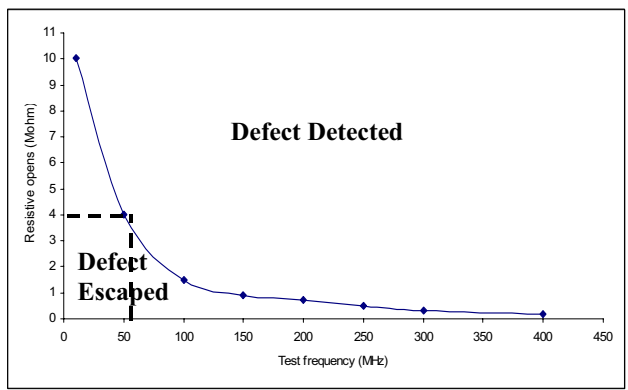

*Figure 8*: Resistive open defect detection vs. test frequency

From our experimental investigation, few devices were selected that failed during at-speed testing. The shmoo plots for those two categories of devices are shown in Figures 9 and 10. The minimum clock period (i.e., maximum frequency) at which the tests are performed is 15ns, which is of course slower compared to the actual speed of the memories (5 to 10 ns). The design limitation of the test chip and the tester hardware setup has great impact on the test frequency. Hence, for our experimental setup, testing the memories at-speed means setting the clock period to 15ns during application of the standard test patterns. This has been achieved by characterizing few fault-free samples while stressing the supply voltage to Vmax (1.95V).

Figure 9 shows the shmoo plot for Chip-3. The characteristic of the shmoo plot clearly reveals that the defect in Chip-3 has lead to a timing failure. Irrespective of the supply voltage the device starts passing after a particular frequency (fail @ 16ns, pass @ 17ns clock period and above).

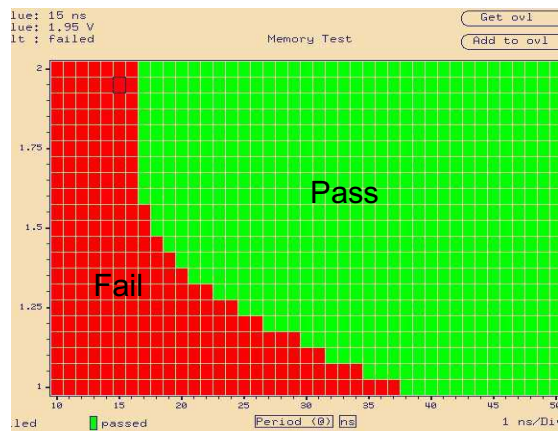

*Figure 9:* Shmoo plot for Chip-3 (Timing failure due to a delay fault)

Figure 10 is another example of an at-speed failure device. The fail characteristic of Chip-4 is different from that of Chip-3, though both lead to timing related failures. In the case of Chip-4, it is to be noted that the delay is also voltage dependent. As the supply voltage is lowered, the pass-fail margin between the faulty chip and fault-free chip reduces; this is a similar observation to what happens when there is a delay fault in random logic. Hence, we may come to the conclusion that the defect in Chip-4 may be present in the periphery of the memory and not in the matrix.

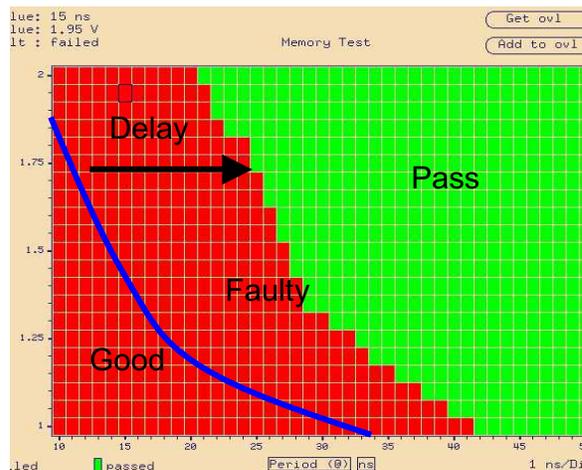

*Figure 10:* Shmoo plot for Chip-4 (Timing failure due to a delay fault)

## 5. Experimental Summary

In this experiment we have tested approximately 11k SRAMs, out of which 36 were selected as interesting devices that passed the standard test but failed the same test patterns at different stress conditions. To be specific, 27 devices failed only VLV testing, 3 devices failed only Vmax and another 3 failed only at-speed test. Additionally, two devices failed both VLV and Vmax testing, and also a single device failed both VLV and at-speed test. Figure 11 shows Venn diagram of the above interesting devices. It is clearly observed from the experimental results that VLV test is the most important stress condition for reducing the test escapes. However, in order to achieve high quality products, the other stress conditions like Vmax and at-speed testing are also necessary.

The other highlight of this investigation is that there is a clear matching between the simulation and the experimental results. As mentioned in Section 3, the Defect Coverage and DPM Estimator has shown a difference of ~9X in DPM level between VLV and Vmax testing, which also can be observed from the experimental data from the Venn diagram as shown in Figure 11.

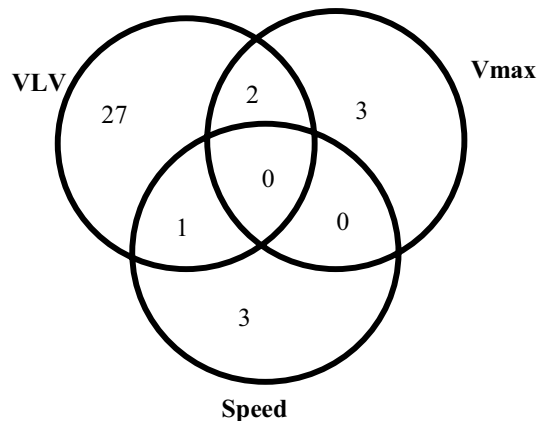

*Figure 11: Venn diagram of failing devices at different stress conditions*

## 6. Conclusions

This paper highlights the importance of the stress conditions for increasing the fault coverage of resistive defects. IFA-based simulation techniques have been implemented to predict the fault coverage and DPM levels at different stress conditions, which helps customers to evaluate the quality of the products upfront.

This investigation shows that VLV testing is the most important stress condition for reducing the DPM level. From our simulation results, we conclude that VLV testing mainly targets resistive bridges while Vmax targets resistive open defects, and finally at-speed test targets timing related failures (dynamic faults) caused by either resistive bridges or opens. The importance of the stress conditions has been validated in an industrial environment.

Test time is an issue during production when we consider the implementation of many algorithms under various stress conditions. Hence, it is recommended to have the best test algorithms combined with specific stress conditions (VLV at low frequency, Vnom and Vmax at high frequency) to reduce test escapes and deliver high quality products.

As continuation of this research, we would like to explore new test algorithms for targeting the soft defects.

Also, physical failure analysis may be carried out to determine the real root cause of these soft defects.

## Acknowledgements
We would like to thank Rutger van Veen, Bram Kruseman, Shaji Krishnan and Rodger Schuttert of Philips Research, for their contributions to this work. This research is partly funded by the European MEDEA Program (TechnoDat) and the memory division of Philips Semiconductors (AMDC).